\documentclass{article}

     \PassOptionsToPackage{numbers, sort&compress}{natbib}

  \usepackage[dblblindworkshop, final]{neurips_2025}
 \workshoptitle{Machine Learning and the Physical Sciences}

\usepackage[utf8]{inputenc} 
\usepackage[T1]{fontenc}    
\usepackage{hyperref}       
\usepackage{url}            
\usepackage{booktabs}       
\usepackage{amsfonts}       
\usepackage{nicefrac}       
\usepackage{microtype}      
\usepackage{xcolor}         

\usepackage{multirow}
\usepackage{wrapfig}
\usepackage{graphicx}
\usepackage{mathtools}
\usepackage{comment}

\usepackage{amsmath}
\usepackage{bm} 

\usepackage{subcaption} 
\usepackage[percent]{overpic}

\title{Self-supervised Synthetic Pretraining for Inference of Stellar Mass Embedded in Dense Gas}

\author{%
  Keiya Hirashima\thanks{RIKEN Special Postdoctoral Researcher} \\
  Center for Interdisciplinary Theoretical\\ and Mathematical Sciences (iTHEMS)\\
  RIKEN \\
  Wako, Japan \\
  \texttt{keiya.hirashima@riken.jp} \\
  \And
   Shingo Nozaki \\
   Department of Earth and Planetary Sciences\\ 
   Kyushu University \\
   Fukuoka, Japan \\
  \texttt{nozaki.shingo.307@s.kyushu-u.ac.jp} \\
  \AND
   Naoto Harada \\
   Department of Astronomy \\
   University of Tokyo \\
   Tokyo, Japan \\
  \texttt{hrdnot-3997@g.ecc.u-tokyo.ac.jp} \\
}

\begin{document}

\maketitle

\begin{abstract}
    Stellar mass is a fundamental quantity that determines the properties and evolution of stars. However, estimating stellar masses in star-forming regions is challenging because young stars are obscured by dense gas and the regions are highly inhomogeneous, making spherical dynamical estimates unreliable. Supervised machine learning could link such complex structures to stellar mass, but it requires large, high-quality labeled datasets from high-resolution magneto-hydrodynamical (MHD) simulations, which are computationally expensive. We address this by pretraining a vision transformer on one million synthetic fractal images using the self-supervised framework DINOv2, and then applying the frozen model to limited high-resolution MHD simulations. Our results demonstrate that synthetic pretraining improves frozen-feature regression stellar mass predictions, with the pretrained model performing slightly better than a supervised model trained on the same limited simulations. Principal component analysis of the extracted features further reveals semantically meaningful structures, suggesting that the model enables unsupervised segmentation of star-forming regions without the need for labeled data or fine-tuning.
\end{abstract}

\section{Introduction}

Stellar mass is a fundamental stellar property that governs luminosity, lifetime, stellar evolutionary tracks, and stellar nucleosynthesis, which produce the chemical elements essential for the origin of our solar system and life. 
In astronomy, the Initial Mass Function (IMF) describes the stellar mass distribution \cite{salpeter1955}. 
Observations suggest that the IMF exhibits a remarkably similar shape across diverse environments,
while the physical mechanisms governing this distribution remain unclear \cite{Offner+2014}.
To uncover the origin of the IMF, it is essential to determine the masses of young, still-forming stars (protostars and pre-main-sequence stars) through observations.
However, determining the masses of young stars from direct observations remains highly challenging.
They are deeply embedded within their natal molecular clouds, making them invisible with optical light.
Furthermore, their luminosity originates primarily from gas accretion rather than stellar radiation, making mass estimates difficult using methods commonly applied to main-sequence stars.

Predicting stellar masses from their environments is a challenging task. The gas is highly inhomogeneous, making analytic models unreliable, while capturing the physics requires three-dimensional (3D) simulations, which are too expensive to produce in large numbers \cite{pelkonen2021}.
A promising approach is to combine high-resolution simulations with deep learning. In our work, three-dimensional (3D) magneto-hydrodynamical (MHD) simulations are employed to capture the physics of star formation and track stellar mass growth \cite{nozaki2025}. 
We then leverage two-dimensional (2D) maps of gas projected from these 3D high-resolution MHD simulations to develop deep learning models for predicting stellar masses. 
Since large amounts of high-resolution simulations or observational data with labels are rarely available, we propose a framework that combines self-supervised pretraining and downstream tasks. Models are trained on numerous synthetic images to learn robust visual representations, while limited high-resolution MHD simulations are reserved for zero-shot and frozen-feature evaluation on downstream tasks.

\section{Related work}
\label{related_work}

\paragraph{Self-supervised Learning for Astrophysics}
Self-supervised learning has become a powerful approach for extracting image representations without labels and could help mitigate challenges from limited labeled data.
Early methods like MoCoV3 \cite{Chen+21} and MAE \cite{He+22} improved vision transformer (ViT) pretraining but often required supervised fine-tuning.
In contrast, DINOv2 \cite{Caron+21, Zhou+21, Oquab+24} captures semantic structures by enforcing consistency across multiple views, transfers to downstream tasks without fine-tuning backbones, and requires only lightweight classifiers or $k$-nearest neighbor evaluation.
In astrophysics, self-supervised learning has enabled galaxy classification with sparse labels \cite{He+21, Desmons+24}, inference of galaxy properties by combining simulations and observations \cite{Eisert+24}, and mitigation of observational biases through metadata \cite{Rizhko+25}, with DINOv2 recently applied to galaxy images \cite{Parker+24}.

\paragraph{Pretraining with Synthetic Data}
Supervised deep learning has achieved remarkable success by training models on large labeled datasets. An alternative line of research investigates the use of synthetic data generated with mathematical equations to achieve competitive performance across various downstream tasks. Such data can be produced inexpensively and in large quantities, without demanding experiments, observations, or extensive computational resources, and without raising ethical or privacy concerns. A pioneering study \cite{Kataoka+20} showed that supervised pretraining on fractal images alone can reach competitive accuracy on natural images, in some cases even surpassing models pretrained on \texttt{ImageNet-22k} \cite{Kataoka+22}. This approach has since been extended to supervised learning with ViTs \cite{Nakashima+22, Kataoka+22, Nakamura+23, Nakamura+24} and to self-supervised learning with convolutional neural networks \cite{Baradad+21, Baradad+22}.


\section{Methodology}

\label{sec:data_gen}
\begin{figure}[ht!]
    \centering
    \includegraphics[width=0.8\linewidth]{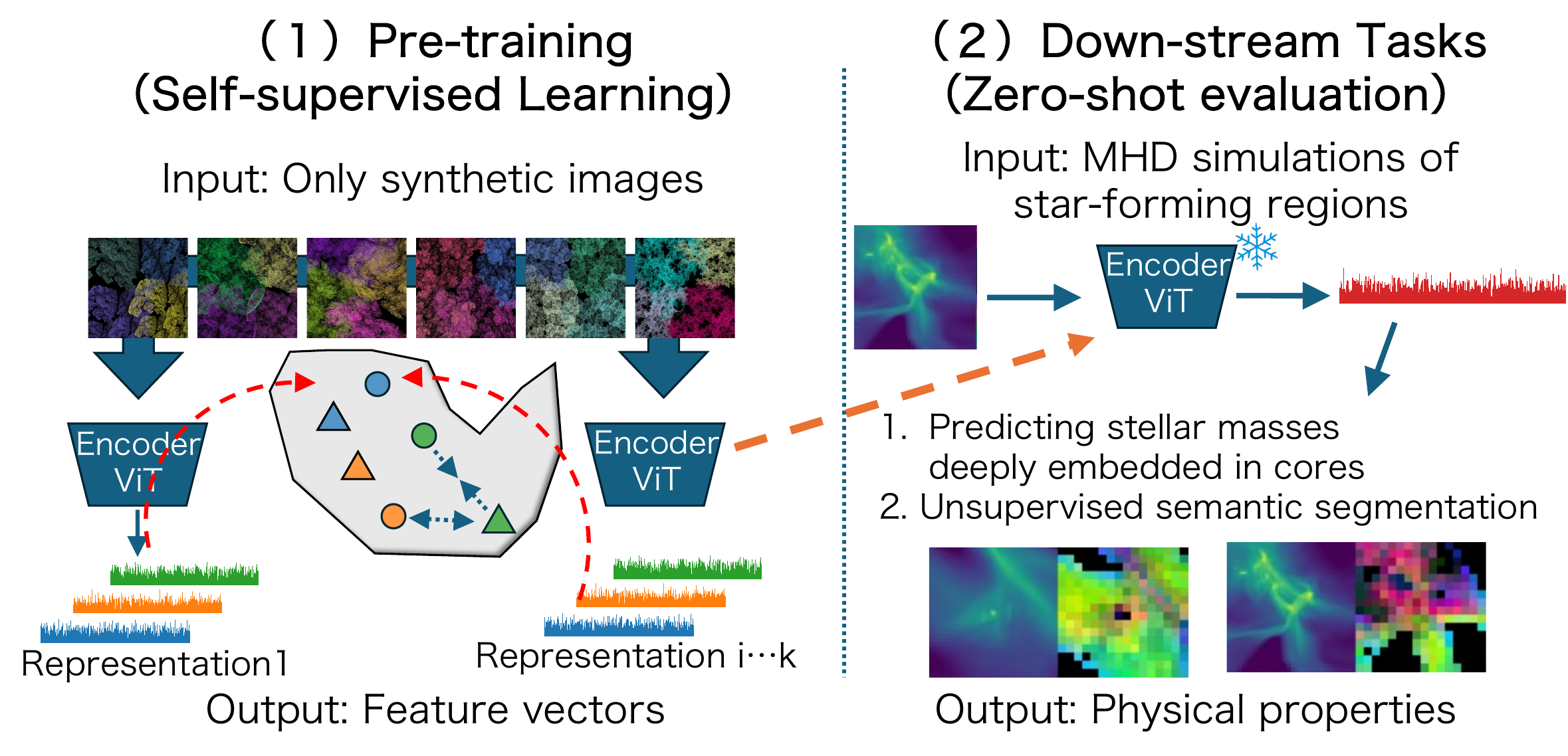}
     \caption{Overview of our model. \textit{Left}: self-supervised pretraining with synthetic fractal images using DINOv2 to extract feature vectors. \textit{Right}: zero-shot evaluation on simulation with the frozen encoder, applied to stellar mass prediction ($k$-NN) and semantic segmentation (PCA-based colors).
}
    \label{fig:model}
\end{figure}

\subsection{Data Generation}

\paragraph{Synthetic Images for Pretraining}
We extend the Flame algorithm \cite{Draves+08} to generate our datasets of fractal images.
With randomly sampled parameters $\theta_i = (a_i, b_i, c_i, d_i, e_i, f_i)$ for rotation and shifting fed to a translation $w$, coordinates are sampled through an iterated function system (IFS) \cite{Barnsley+88}, 
\begin{equation}
w (\bm{x}; \theta_i)
=
\begin{pmatrix}
  a_i & b_i \\
  c_i & d_i
\end{pmatrix}
\bm{x}
+
\begin{pmatrix}
  e_i \\
  f_i
\end{pmatrix},
\end{equation}
where $\bm{x}$ is a coordinate.
At each sampling step, one of the non-linear variations (e.g., spherical and bubble) of the original Flame algorithm \cite{Draves+08} is probabilistically applied, yielding the next sampled point $\bm{x}_{i+1} = w (\bm{x}_i; \theta_i)$.
Those sampled points are interpolated to $336 \times 336$ images.
We draw candidate frames with one million points, accept only those that cover $\ge 80\%$ of the image plane, and thus build a dataset of 1M images.
Examples are shown in Fig.~\ref{fig:model} (1).
The dataset generation was executed with a throughput of 2.67 TFLOPS per image and 2.67 EFLOPS in total.

\paragraph{Simulations of star-forming regions}
We performed 3D MHD simulations with \texttt{SFUMATO}, an adaptive mesh refinement code \cite{matsumoto2007,matsumoto2015,fukushima2021,nozaki2025}, in a cubic box of 4 parsec\footnote{1 parsec $\sim 3.08 \times 10^{13}$ km $\sim$ 3.26 light years} per side containing 3000 M$_\odot$\footnote{1 M$_\odot$ (solar mass) is equal to the mass of the Sun.} of gas with an initial uniform proton density of $1365~\mathrm{cm^{-3}}$ and a magnetic field of $10\,\mu\mathrm{G}$ along the $z$-axis.
To follow the long-term evolution, we use the sink particle method \cite{matsumoto2015}, in which unstable dense clumps are replaced by sink particles that accrete gas within a fixed radius ($5.0 \times 10^{-4}$ pc).
The accreted mass is taken as the protostellar mass, allowing us to trace protostellar growth. The finest spatial resolution is $\Delta x \sim 3 \times 10^{-3}$ parsec, sufficient to resolve the Jeans length with more than five cells, given an initial velocity field with a Mach number of 10. Our dataset consists of 32k snapshots of 0.5 pc regions centered on protostars, from which we construct $64 \times 64$ maps of column density, mean line-of-sight velocity, and its velocity dispersion along the $x$, $y$, and $z$ axes, each paired with the elapsed time since protostar formation and the stellar mass shown in Fig.~\ref{fig:model} (2). 
The simulation was executed with a throughput of 2540 TFLOPS per snapshot and 81.2 EFLOPS in total.

\subsection{Model Implementation}
We employ a ViT-L/16 encoder within the DINOv2 framework for self-supervised pretraining and zero-shot or frozen-feature evaluation. The encoder is pretrained on 1M synthetic fractal images at a resolution of 336 for 100 epochs with a batch size of 1024 and a patch size of 16. For comparison, we implement a ResNet-18 \cite{resnet} baseline trained in a fully supervised manner. Both models use a cosine-annealed learning rate schedule with a maximum of 0.04, including 10 warm-up epochs followed by 90 epochs of cosine decay. The ResNet-18 is trained with an $L_2$ regression loss. Simulation data are preprocessed by applying a logarithmic transformation to stellar mass and column density, and min–max normalization to mean line-of-sight velocity and its dispersion.

\subsection{Experiments}

\paragraph{Self-supervised Pretraining with Synthetic Data and $k$-NN Regression} 
The pretrained ViT-L/16 encoder is applied to 32k snapshots from star-formation simulations at a resolution of 64 to obtain 1024-dimensional feature vectors. Principal component analysis (PCA) is fitted on the training split and applied to all features while preserving the full dimensionality of 1024 (PCA whitening). The transformed features are then evaluated with a distance-weighted $k$-nearest neighbors ($k$-NN) regressor ($k=5$) to predict the logarithm of stellar mass, using 24k training and 8k test samples. Prediction performance is assessed in terms of root-mean-square error (RMSE) and the coefficient of determination ($R^2$). 

\paragraph{Zero-shot Unsupervised Feature Visualization} 
To examine the semantic structure of the learned representations, feature vectors from the pretrained ViT-L/16 encoder are projected with PCA and the first three components are mapped to the RGB color space. Although pretraining is performed with a patch size of 16, for visualization we linearly upsample $4{\times}4$ input patches to $16{\times}16$ prior to encoding, in order to maintain consistency with the token granularity of the model.

\section{Results}
\label{sec:results}

\begin{figure}[ht!]
  \centering
  \begin{subfigure}[t]{0.33\linewidth}
    \includegraphics[width=\linewidth]{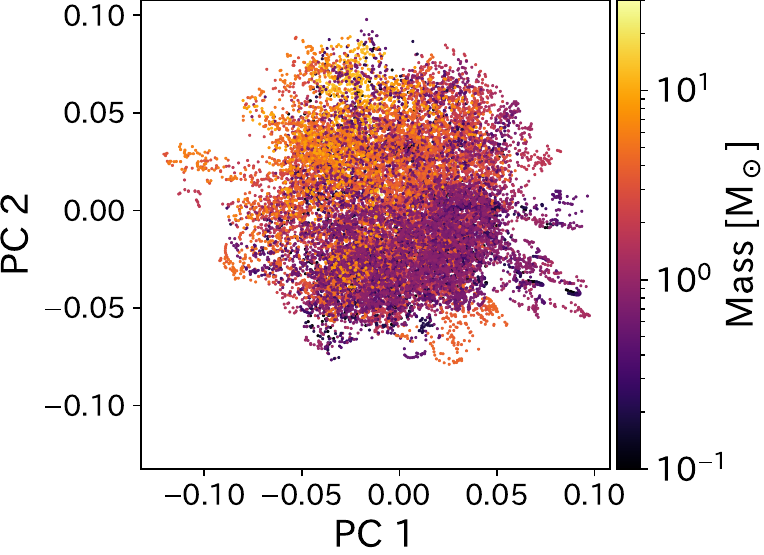}
    \caption{}
    \label{fig:pca_dino}
  \end{subfigure}\hfill
  \begin{subfigure}[t]{0.32\linewidth}
    \includegraphics[width=\linewidth]{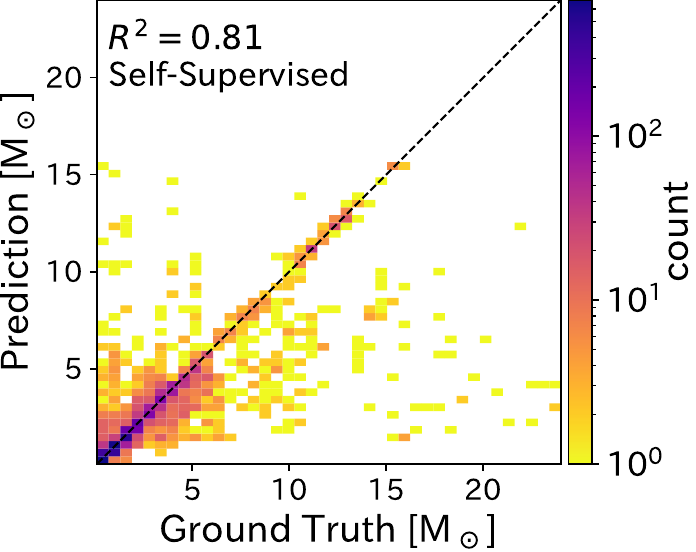}
    \caption{}
    \label{fig:r2_DINO}
  \end{subfigure}\hfill
  \begin{subfigure}[t]{0.32\linewidth}
    \includegraphics[width=\linewidth]{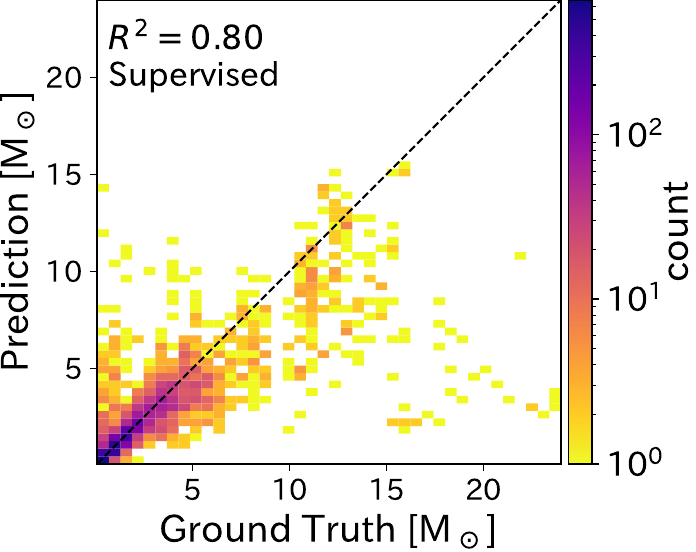}
    \caption{}
    \label{fig:r2_resnet}
  \end{subfigure}

  \caption{Frozen-feature regression of stellar masses. 
\textbf{(a)} PCA projection of feature vectors from DINOv2 colored by stellar mass. 
\textbf{(b)} True versus predicted stellar masses using DINOv2 representations with $k$-NN regression. 
\textbf{(c)}True versus predicted stellar masses from a supervised ResNet-18 baseline.
}
  \label{fig:three}
\end{figure}

We evaluate the ViT-L/16 encoder pretrained with fractal images on zero-shot and frozen-feature tasks, keeping the model parameters frozen.

\paragraph{Frozen-feature Regression on Stellar Masses}
Fig.~\ref{fig:pca_dino} shows a scatter plot of the first and second PCA components of feature vectors from column density, mean line-of-sight velocity, and its velocity dispersion maps.
The distribution exhibits a weak trend with stellar masses, which are otherwise difficult to infer from 2D density and velocity information alone.  
To evaluate predictive performance, we use all PCA components with a $k$-NN regressor trained on the training set and tested on the validation set. Fig.~\ref{fig:r2_DINO} and Fig.~\ref{fig:r2_resnet} compare stellar mass predictions from our method and a supervised ResNet-18 baseline, respectively. Both approaches follow the ground truth trend up to $\sim$ 6 M$_\odot$, where more than $10^2$ training samples are available. Beyond this, with fewer than 10 samples, ResNet-18 tends to underestimate stellar masses, while DINOv2 captures many true values in the range 6--15 M$_\odot$. At higher masses, neither model performs reliably due to data scarcity. Table~\ref{tab:supervised_selfsupervised_model} shows that PCA features slightly improve scores over raw feature vectors and that pretraining with fractal images markedly improves performance compared to the result of random initialization with DINOv2.

\begin{table}[h]
    \centering
    \begin{tabular}{l|cc|ccc}
               &   \multicolumn{2}{c|}{ResNet-18} &  \multicolumn{3}{c}{DINOv2 + $k$-NN (k=5)}\\ 
         Methods &  Random Init.  & Pretrained  & Random Init.  & Pretrained & with PCA whitening\\ \hline
         $R^2 (\uparrow)$    & -1.9  & 0.80 &    -0.58 & 0.80 & \textbf{0.81}\\
         RMSE $(\downarrow)$    & 0.34  & 0.089    & 0.52 & 0.089 & \textbf{0.088}       
    \end{tabular}
    \vspace{5pt}
    \caption{$R^2$ and RMSE of frozen-feature regression on stellar mass using ResNet-18 and DINOv2.}
    \label{tab:supervised_selfsupervised_model}
\end{table}



\paragraph{Zero-shot Semantic Segmentation with PCA-based Colors}
\begin{figure}[ht!]
  \centering
  \begin{subfigure}[t]{0.48\linewidth}
    \begin{overpic}[width=\linewidth]{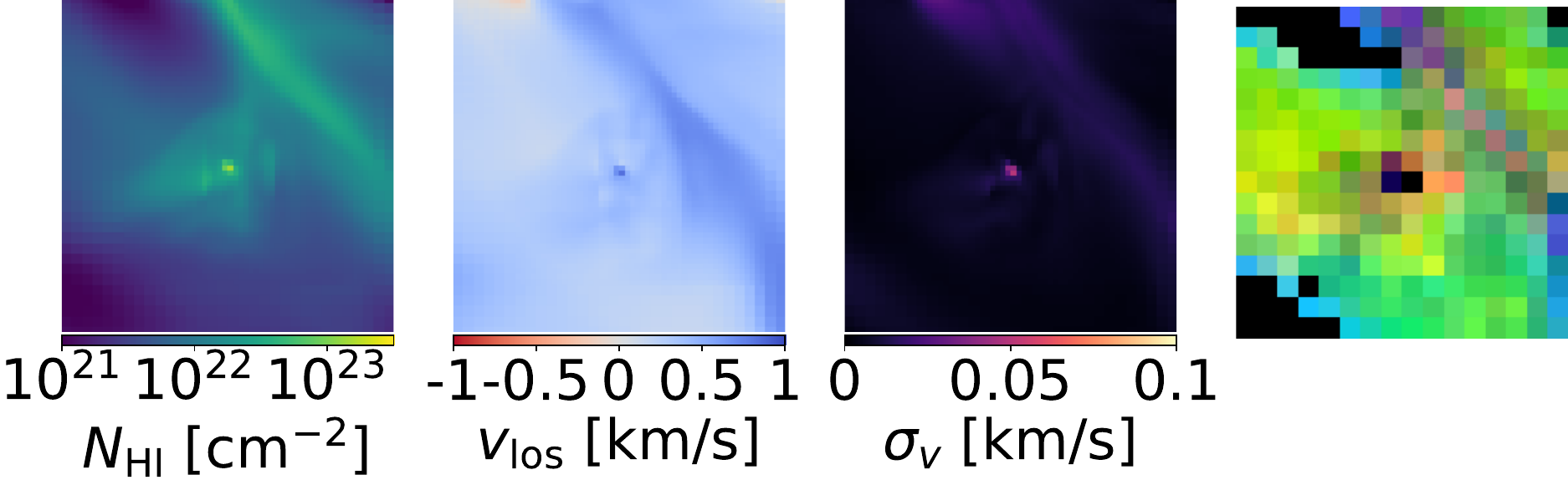}
      \put(-3,25){\textbf{(a)}} 
    \end{overpic}
    \phantomsubcaption\label{fig:ex-a}
  \end{subfigure}\hfill
  \begin{subfigure}[t]{0.48\linewidth}
    \begin{overpic}[width=\linewidth]{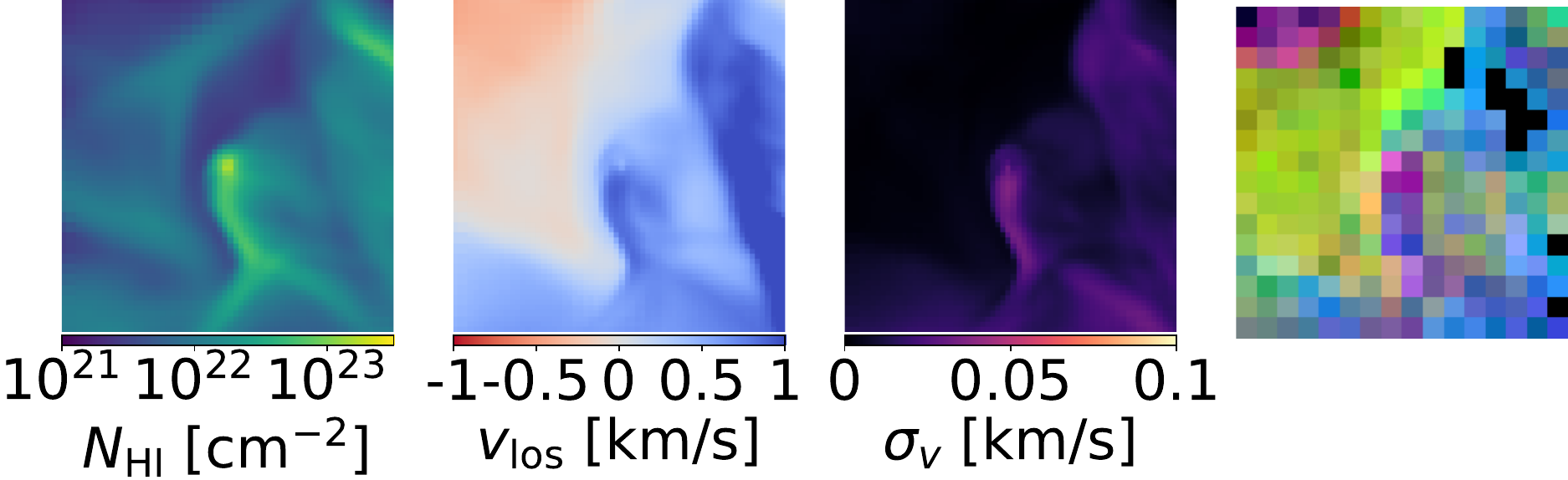}
      \put(-3,25){\textbf{(b)}}
    \end{overpic}
    \phantomsubcaption\label{fig:ex-b}
  \end{subfigure}
  \vspace{-0.3cm}
  \begin{subfigure}[t]{0.48\linewidth}
    \begin{overpic}[width=\linewidth]{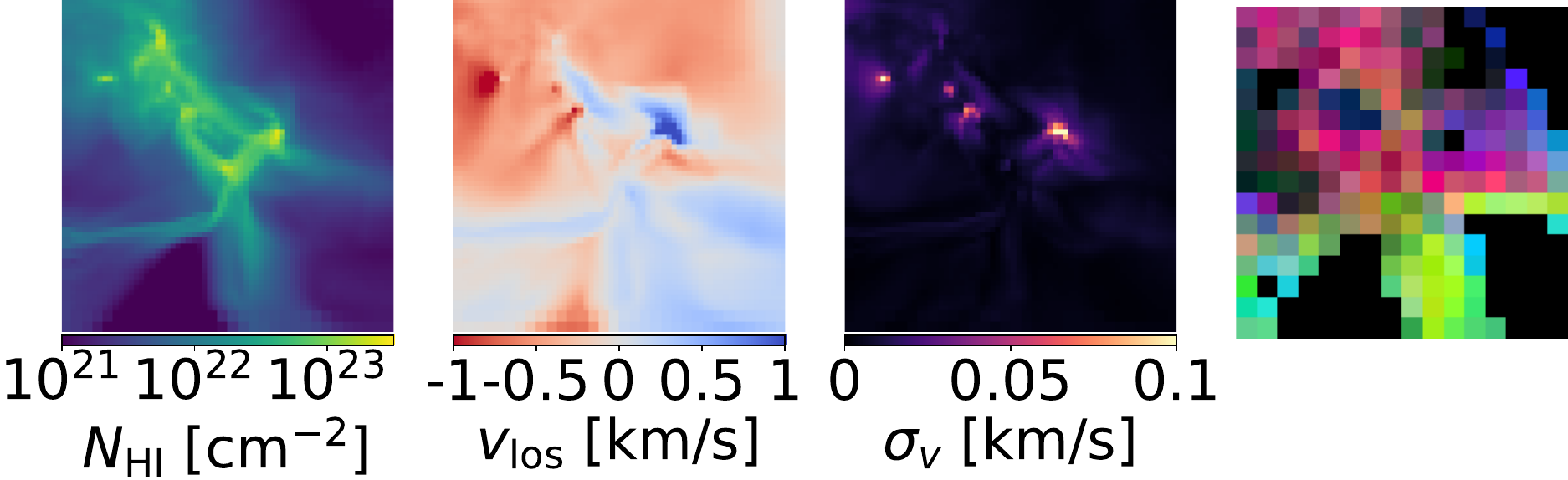}
      \put(-3,25){\textbf{(c)}}
    \end{overpic}
    \phantomsubcaption\label{fig:ex-c}
  \end{subfigure}\hfill
  \begin{subfigure}[t]{0.48\linewidth}
    \begin{overpic}[width=\linewidth]{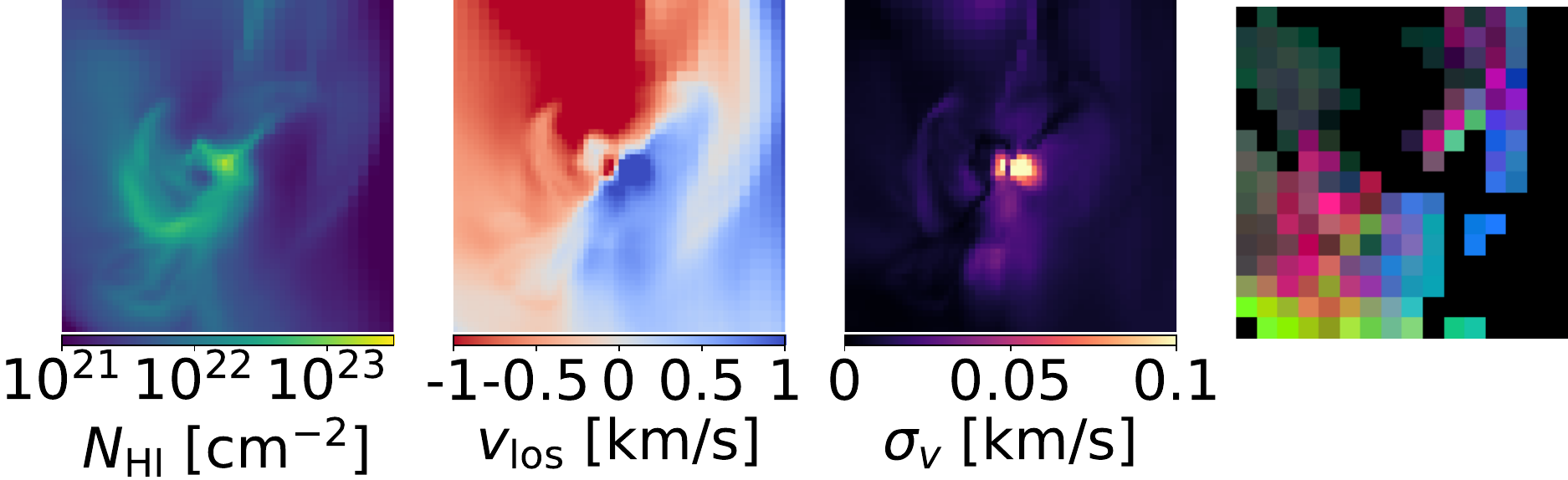}
      \put(-3,25){\textbf{(d)}}
    \end{overpic}
    \phantomsubcaption\label{fig:ex-d}
  \end{subfigure}\hfill
  \caption{Snapshots from MHD simulations with visualizations of PCA components of feature vectors. Each panel shows four maps: column density $N_\mathrm{HI}$, mean line-of-sight velocity $v_\mathrm{los}$, its velocity dispersion $\sigma_v$, and a color map of the first three PCA components from image patches.
}
  \label{fig:semantic}
\end{figure}

Fig.~\ref{fig:semantic} shows four examples (a–d), each containing four panels: column density $N_\mathrm{HI}$, mean line-of-sight velocity $v_\mathrm{los}$, its velocity dispersion $\sigma_v$, and a color map based on the first three PCA components of the 1024-dimensional feature vectors. 
Black areas correspond to either diffuse, low-density regions or regions of very high velocity dispersion, the latter likely marking sites of ongoing star formation.  
Yellow to yellow-green areas highlight regions of low velocity dispersion (Fig.~\ref{fig:ex-a}--\ref{fig:ex-c}).  
Magenta and dodgerblue indicate negative and positive line-of-sight velocities in regions of high velocity dispersion (Fig.~\ref{fig:ex-b}--\ref{fig:ex-d}), where gas may accrete onto dense cores and contribute to stellar growth.  
Notably, this semantic segmentation arises directly from the PCA projection of pretrained representations, without any labeled data or supervised fine-tuning.

\section{Summary and Discussion}


Our results demonstrate that self-supervised synthetic pretraining can serve as a data-efficient alternative to supervised pipelines in high-resolution, yet data-limited, MHD simulations, with a ViT encoder achieving performance comparable to that of supervised learning. PCA-based visualization further revealed meaningful structures, such as dense cores and inflows, motivating the extension of this approach to stellar property prediction and its direct application to observational data.

The present approach indicates a potential to predict protostellar masses, and incorporating information from more extended gas and velocity fields may ultimately enable predictions of the final stellar mass and the IMF. The framework, however, still relies on labeled simulation data for training in order to apply it to observations. Meanwhile, PCA-based segmentation highlights that broad structural patterns can be identified without labels, though addressing observational noise—potentially by constructing datasets from the noise itself—will be crucial for robustness in practice.


\medskip

\begin{ack}
The authors thank Takaharu Yaguchi, Kazuki Tokuda, Yoshito Shimajiri, and Kana Moriwaki for fruitful discussions.
This work was initiated following discussions at the International Conference on Scientific Computing and Machine Learning 2025 and the 37th Rironkon (Community of Theoretical Astrophysics) annual symposium 2024.
KH is supported by the Special Postdoctoral Researchers Program at RIKEN.
SN is supported by JST SPRING, Grant Number JPMJSP2136.
NH is supported by JSPS KAKENHI Grant Number JP25K17434.
Numerical computations were carried out on Flatiron Institute’s Rusty computing cluster, the Cray XC50 (Aterui II) at the Center for Computational Astrophysics of the National Astronomical Observatory of Japan, and the Yukawa-21 at the Yukawa Institute for Theoretical Physics (YITP), Kyoto University.
We gratefully acknowledge the Flatiron Institute and the Scientific
Computing Core for their support.
\end{ack}

{
\small
\bibliographystyle{nips}
\bibliography{neurips25}
}





\newpage

\end{document}